\documentclass[twocolumn]{aastex62}

\newcommand{\Ni}{\ensuremath{^{56}\mathrm{Ni}}}

\newcommand{\Msun}{\ensuremath{\mathrm{M}_\odot}}
\newcommand{\Rsun}{\ensuremath{\mathrm{R}_\odot}}

\newcommand{\tdelay}{\ensuremath{t_\mathrm{delay}}}
\newcommand{\tcut}{\ensuremath{t_\mathrm{cut}}}

\newcommand{\Msunps}{\ensuremath{\Msun~\mathrm{s^{-1}}}}

\usepackage{ulem}

\graphicspath{{./}{figures/}}

\received{March 4, 2019}
\revised{May 27, 2019}
\accepted{May 31, 2019}

\shorttitle{Fallback accretion powered SNe}
\shortauthors{Moriya et al.}

\begin{document}

\title{
Fallback accretion powered supernova light curves based on a neutrino-driven explosion simulation of a 40~\Msun\ star
}

\correspondingauthor{Takashi J. Moriya}
\email{takashi.moriya@nao.ac.jp}

\author[0000-0003-1169-1954]{Takashi J. Moriya}
\affil{Division of Science, National Astronomical Observatory of Japan, National Institutes of Natural Sciences, 2-21-1 Osawa, Mitaka, Tokyo 181-8588, Japan}

\author{Bernhard M\"uller}
\affil{Monash Centre for Astrophysics, School of Physics and Astronomy, Monash University, VIC 3800, Australia}
\affil{Astrophysics Research Centre, School of Mathematics and Physics, Queen’s University Belfast, Belfast BT7 1NN, UK}

\author{Conrad Chan}
\affil{Monash Centre for Astrophysics, School of Physics and Astronomy, Monash University, VIC 3800, Australia}

\author[0000-0002-3684-1325]{Alexander Heger}
\affil{Monash Centre for Astrophysics, School of Physics and Astronomy, Monash University, VIC 3800, Australia}
\affil{Tsung-Dao Lee Institute, Shanghai 200240, China}

\author{Sergei I. Blinnikov}
\affil{National Research Center "Kurchatov institute", Institute for Theoretical and Experimental Physics (ITEP), 117218 Moscow, Russia}
\affil{Kavli Institute for the Physics and Mathematics of the Universe (WPI), The University of Tokyo Institutes for Advanced Study, The University of Tokyo, 5-1-5 Kashiwanoha, Kashiwa, Chiba 277-8583, Japan}
\affil{Sternberg Astronomical Institute, M.V. Lomonosov Moscow State University, Universitetski pr. 13, 119234 Moscow, Russia}









\begin{abstract}
We present synthetic light curves of fallback-powered supernovae based on a neutrino-driven explosion of a 40~\Msun\ zero-metallicity star
with significant fallback accretion onto a black hole that was
previously simulated by \citet{chan2018bh} until shock breakout.  Here, we investigate the light curve properties of the explosion after shock breakout
for various fallback models.   Without extra power from fallback accretion, the 
light curve is that of a Type~IIP supernova with a plateau magnitude 
of around $-14~\mathrm{mag}$ and a
plateau duration of 40~days. With extra power for the light curve
from fallback accretion, however, we find that the transient can be significantly more luminous. The light-curve shape can be SN~1987A-like or Type~IIP-like, depending on the efficiency of
the fallback engine. If the accretion disk forms soon after the collapse and more than 1\% of the accretion energy is released as the central engine,
fallback accretion powered supernovae become as luminous as superluminous supernovae.  We suggest that Type~II superluminous supernovae with broad hydrogen features could be related to such hydrogen-rich supernovae powered by fallback accretion.  In the future, such superluminous supernovae powered by fallback accretion might be found among the supernovae from the first stars in addition to pair-instability supernovae and pulstational pair-instability supernovae. 
\end{abstract}



\section{Introduction}\label{sec:introduction}
The core collapse of massive stars is the major formation site of stellar mass black holes (BHs). 
The collapse initially leads to the formation of a proto-neutron star and the formation of a shock wave after the core overshoots nuclear density and rebounds. The shock quickly stalls, however, and the proto-neutron star continues to grow by accretion.
In many cases, the shock is then revived, most likely
due to heating by neutrinos which are emitted
in copious amounts from the proto-neutron star  \citep[e.g.,][]{muller2016cc}, and a successful supernova
(SN) appears.
In some progenitors, shock revival may instead be achieved
by magnetohydrodynamic effects
\citep[e.g.,][]{akiyama2003mri,moesta2015mhd} or other mechanisms \citep[e.g.,][]{fischer2018blinnikov}.
If the shock is not revived, however, the outer shells
of the star keep collapsing and a BH is eventually formed.
There may be a transition regime, however, in which both a BH is formed
and an observable SN transient is produced.
It has been proposed that SNe with small explosion energies may still be accompanied by BH formation \citep{zampieri2003bhsn,moriya2010fallback}, and such SNe are suggested to be important to account for peculiar chemical abundances in low-metallicity stars \citep[e.g.,][]{keller2014pop3fb,bessell2015pop3fb,ishigaki2014pop3fb}.
Clarifying the  pathways to BH formation and their accompanying transients is important, especially for understanding the origin of the massive double BH binaries recently observed with gravitational waves \citep[e.g.,][]{abbott2016bhbh}.

Several scenarios for transients from BH-forming events
have been considered in the literature.
Even if the shock is not revived and a BH forms directly
by ongoing accretion, the reduction of the star's gravitational
mass by neutrino emission can still trigger
mass ejection from the progenitor surface, which then gives
rise to a faint and slowly evolving transient
\citep{nadezhin1980,lovegrove2013failed,fernandez2018failed,coughlin2018failed}. 

Another possibility is that a central engine operates
\emph{after} BH formation and initiates an explosion.
For example, in the collapsar scenario for gamma-ray bursts the collapse may form an accretion disk,
from which relativistic jets can be launched by magnetohydrodynamic effects
or neutrinos \citep[e.g.,][]{woosley1993col,macfadyen1999col},
while the disk wind powers the accompanying supernova \citep[e.g.,][]{macfadyen1999col,hayakawa2018maeda}.

Such an accretion-powered engine could also operate if an explosion has already been triggered
by another mechanism (e.g., by neutrino heating), but is not sufficiently energetic to unbind the whole star.
In this case, some of the ejecta eventually fall back to the central remnant \citep[e.g.,][]{chevalier1989fallback,zhang2008fallback}.
Similar to the collapsar scenario,
such long-term fallback accretion could lead to the formation of an accretion disk 
and the launching of an accretion disk wind (\citealt{dexter2013fallback,feng2018fallback}, see also \citealt{HW12,gilkis2014jet}). The accretion disk wind collides with the accreting material and forms strong shocks that can push back some of the matter undergoing fallback. Moreover, the shocks provide a power source to make the ejecta bright \citep{dexter2013fallback}. Such fallback-powered SNe have been related to peculiar SNe such as superluminous SNe (SLSNe; \citealt{dexter2013fallback,moriya2018fallbackslsn}).  In particular, \citet{moriya2018fallback} suggested that the energetic hydrogen-rich SN OGLE-2014-SN-073 \citep{terreran2017ogle73} can be explained as a fallback-powered SN.

It is challenging, however, to model fallback-powered SNe consistently.
In a previous study of fallback-powered SNe \citep{moriya2018fallback}, we 
used a semi-analytic method to estimate the shock velocity propagating in SN progenitors to derive the fallback accretion rate onto the central BH. The initial explosion was triggered artificially in a spherically symmetric code, and the explosion energy was chosen by hand.
Although a number of multi-dimensional simulation of fallback in BH forming SNe have been conducted
in recent years
\citep{joggerst2009mixing,joggerst2010mixing,joggerst2010mixingnuc,chen2017popiii} and
could be used as input for light curve (LC) calculations,
these were also artificially initiated by a thermal bomb or a kinetic piston instead
of modelling the phase of shock revival self-consistently.
Recently, \citet{chan2018bh} simulated a neutrino-driven fallback SN of a massive (40~\Msun) zero-metallicity progenitor from core collapse to shock breakout in three dimensions (3D) for the first time. The explosion of this progenitor has been suggested to explain the chemical properties of the recently discovered iron-free star \citep{bessell2015pop3fb}. 
Different from previous studies, \citet{chan2018bh} find that the transfer of energy from the core of
the SN to the ejected shells is inherently multi-dimensional during the first tens of seconds
after black hole formation, although the explosion has become nearly spherical at shock breakout.
In this study, we investigate the LC properties of the fallback-powered SNe based on this state-of-the-art simulation presented by \citet{chan2018bh} as a step towards
more consistent modelling of fallback-powered transients.

This paper is organized as follows: We present the model setup of our LC calculations in Section~\ref{sec:setup}. The results of our LC calculations are shown in Section~\ref{sec:lightcurve}. We discuss our results and conclude this paper in Section~\ref{sec:discussion}.

\section{Model setup}\label{sec:setup}
\subsection{Explosion model}
We adopt the  3D fallback SN model of \citet{chan2018bh} as input
for our LC calculations in this study.  
The SN progenitor is 
a 40~\Msun\ mass star of zero metallicity and was evolved using
the stellar evolution code \texttt{Kepler} up to the onset
of collapse \citep{heger2010zmstar}.  The progenitor has a hydrogen-rich envelope of 24.7~\Msun\ and a helium core mass of 15.3~\Msun. Its radius is 24~\Rsun. At the onset of collapse, the progenitor is mapped to the
relativistic neutrino hydrodynamics code \texttt{CoCoNuT-FMT},
and evolved in 3D through bounce, shock revival, and into the explosion phase until BH formation. At the time of BH formation, the hydrodynamic structure is mapped into the 
quasi-Lagrangian moving-mesh hydrodynamics code
\texttt{Arepo}  \citep{springel2010arepo}, and the hydrodynamic evolution is followed until shock breakout, which occurs 3900~sec after the onset of the core collapse. For further details, we refer to
\citet{chan2018bh}.

At shock breakout, we spherically average the model to obtain initial
conditions for our LC calculations. 
Since the strongly asymmetric neutrino-heated ejecta have already undergone
fallback and the 3D model has a nearly spherically symmetric structure 
at the time of shock breakout, the assumption of the spherical symmetry 
is reasonable for the subsequent evolution \citep{chan2018bh}. 
At this stage,  the outer layers have a positive explosion energy of $\sim 2\times 10^{50}~\mathrm{erg}$, and significant fraction of the progenitor is ejected without the
need to assume any additional energy input from a central
BH engine. The central BH mass at the time of shock breakout is 19~\Msun. The bound and unbound ejecta masses are 10~\Msun\ and 11~\Msun, respectively, and the final BH mass is estimate to be 29~\Msun.

\subsection{Fallback accretion rate}
Based on the spherically averaged hydrodynamic structure of the model at shock breakout,
we estimate the fallback accretion rate onto the central BH.
We follow the subsequent hydrodynamic evolution of the fallback material
using the ballistic approximation, i.e., we only take into account the effect of gravity but neglect pressure forces. In other words, we simply solve the equation of motion $d^2r/dt^2 = -G M(r)/r^2$ with the initial conditions at shock breakout, where $M(r)$ is the mass coordinate.
Using this assumption, we can simply take the initial velocity and the inner mass to estimate the fallback accretion rate. The accretion rate at each time is estimated at a radius $10^8~\mathrm{cm}$.

\begin{figure}
\epsscale{1.2}
\plotone{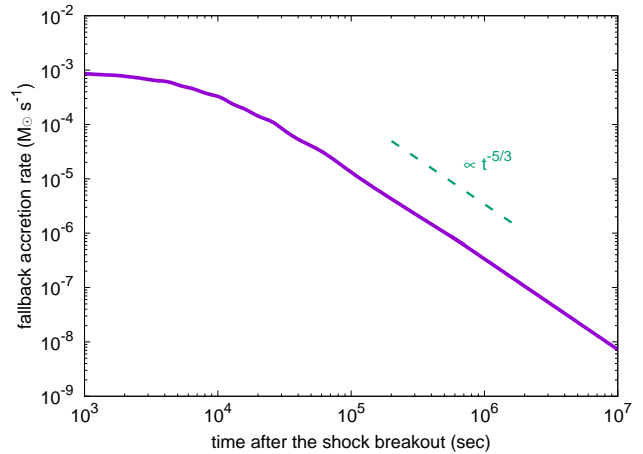}
\caption{Fallback accretion rate estimated from the neutrino-driven explosion model of \citet{chan2018bh}. \label{fig:accrate}}
\end{figure}

Figure~\ref{fig:accrate} shows the estimated fallback accretion rate. Until shock breakout,
the BH accretion rate can also be obtained from  the \texttt{Arepo} simulation
for a consistency check: it is $9\times 10^{-4}~\Msun~\mathrm{s^{-1}}$ shortly before the shock breakout.  This is consistent with and validates the simplified estimate from the ballistic approximation. The fallback accretion rate is of the order of $10^{-4}~\Msunps$ in the first $10^{4}$~sec. Later, the accretion rate becomes proportional to $t^{-\frac{5}{3}}$, as is expected from analytic theory \citep{michel1988fallback,chevalier1989fallback}.  The overall evolution of the accretion rate is consistent with other analytic estimates \citep[e.g.,][]{dexter2013fallback}.

\subsection{Light-curve calculations}
We follow the LC evolution of the explosion model using the one-dimensional multi-group radiation hydrodynamics code \texttt{STELLA} \citep{blinnikov1998sn1993j,blinnikov2000sn1987a,blinnikov2006sniadeflg},
which we already used in our previous study of fallback-powered SNe \citep{moriya2018fallback}.
Briefly, \texttt{STELLA} calculates the spectral energy distributions (SEDs) at each time-step and we can obtain multicolour LCs by convolving filter functions with the SEDs. In addition to the hydrodynamic equations, \texttt{STELLA} implicitly treats time-dependent equations of the angular moments of intensity averaged over a frequency bin with the variable Eddington method. Our calculations are performed by adopting 100 frequency bins from 1~\AA\ to 50000~\AA, which is standard in \texttt{STELLA}. Local thermodynamic equilibrium is assumed to set the ionization levels of materials determining opacity.

We take both the hydrodynamic structure and the chemical abundances from
the spherically-averaged 3D model at shock breakout.
The abundances in the ejecta are basically the same as that of the initial metal-free primordial gas, because \citet{chan2018bh} found no significant global mixing in the ejecta during the explosion. This also justifies our use of the spherically averaged abundance in our LC calculations.

We inject the fallback accretion power close to the inner boundary of the \texttt{STELLA} model in the form of thermal energy.
Thermal energy is injected in 0.1~\Msun\ above the mass cut at $3\times 10^{11}~\mathrm{cm}$.
This mass cut is chosen such that our \texttt{STELLA} calculations can be performed smoothly. Although the disk formation and the actual energy injection could occur at smaller radii, this difference is not likely to affect LCs during the epochs we mainly discuss (after $\sim 1~\mathrm{day}$). This is because photosphere is well above this injection radius.
The thermal energy injection rate, $L_\mathrm{fallback}$, follows
\begin{equation}
    L_\mathrm{fallback} = \eta \dot{M}c^2,
\end{equation}
where $\dot{M}$ is the fallback accretion rate (Fig.~\ref{fig:accrate}), and $c$ is the speed of light. $\eta$ is the conversion efficiency from fallback accretion to the central energy input.
The mechanism of energy transfer from
fallback energy to ejecta, in which different physical mechanisms may be at work, is not obvious. Such a mechanism is beyond the scope of our current paper and therefore we simply parametrized it by introducing the efficiency $\eta$. For example,
the conversion efficiency is estimated to be $\eta \sim 10^{-3}$ by \citet{dexter2013fallback}, but is in fact quite uncertain. We conduct our LC calculations by varying $\eta$ between $10^{-2}$ and $10^{-4}$. Because most of the {\Ni} made by the SN shock has already been accreted to the central remnant at the time of the shock breakout, we ignore \Ni\ heating in this study.

We assume several delay time ($t_\mathrm{delay}$) models for the fallback energy input to be activated. 
Delayed energy injection could be caused, e.g., by a delay in forming the fallback accretion disk.  Whereas the ``no delay'' model ($\tdelay = 0$) has central fallback accretion power right from the beginning of the LC calculations, $\tdelay = 10^{4}~\mathrm{sec}$, $10^{5}~\mathrm{sec}$, and $10^6~\mathrm{sec}$ models have a delay of the central energy input by $10^4$~sec, $10^5$~sec, and $10^{6}$~sec, respectively.
The delay time is determined by the remaining angular momentum in the progenitor and the imparted angular momentum during the explosion to form the accretion disk and we simply assume several possibilities in this study.
Progenitor rotation
was not included in \citet{chan2018bh}, but stellar evolution models that include
angular momentum transport by the Tayler-Spruit dyanmo typically have sufficient
angular momentum for disk formation in the large parts of the hydrogen shell if
mass loss is weak as for Population~III stars, even if they rotate subcritically
on the zero-age main sequence (see, e.g., the 30~\Msun\ model m30vk02 of
\citealt{yoon2012popiii}). In the model of \citet{chan2018bh}, the accretion
of the hydrogen shell on the BH roughly starts around shock breakout
at $\sim4000~\mathrm{sec}$ (see their Figure~4), so it is realistic to expect a delay time of the order of $10^4~\mathrm{sec}$ or longer
until the fallback engine can operate in the case of moderately fast progenitor
rotation.
Although the angular momentum left in the original explosion model of \citet{chan2018bh} is rather small and \tdelay\ could be even larger, the angular momentum can be gained by some ways as discussed in Section~\ref{sec:discussion}. Therefore, we simply parameterize \tdelay\ and investigate several possibilities. We show the results of $\tdelay\lesssim 10^{6}~\mathrm{sec}$ because the injected energy is mostly less than the original explosion energy of $\sim 2\times 10^{50}~\mathrm{erg}$ when $\tdelay\gtrsim 10^7~\mathrm{sec}$ and we do not expect a significant effect on LCs by the fallback energy when $\tdelay\gtrsim 10^{7}~\mathrm{sec}$. Even for the case of $\tdelay=10^{6}~\mathrm{sec}$, the $\eta=10^{-4}$ model does not obtain enough energy to affect the LC.
The models with $\tdelay \lesssim 10^3~\mathrm{sec}$ are found to be similar to those without a delay and the delay times larger than $10^3~\mathrm{sec}$ makes a difference in LCs.
We also show the LC without any central energy input, assuming that the fallback material is simply swallowed by the central BH. 

It is possible that the BH accretion power stops at some time, e.g., because the outflows powered
by fallback accretion early on
obstruct the fallback of shells further out. In order to see the effect of a cut of fallback accretion power, we 
introduce a parameter  \tcut\ for the cut-off time of the fallback-powered engine. We set $\tcut = \infty$, i.e., no power cut, in our standard simulations, but we also present LCs with $\tcut = 10$, $50$, $100$, $150$, and $200$~days to show the effect of variations in \tcut.
In this work, we are interested in the effect of \tcut\ in late phases after the successful energy injection. The quantities \tcut\ and \tdelay\ may actually be similar, but such a case is beyond the scope our paper.

\section{Light curves}\label{sec:lightcurve}
\begin{figure}
\epsscale{1.2}
\plotone{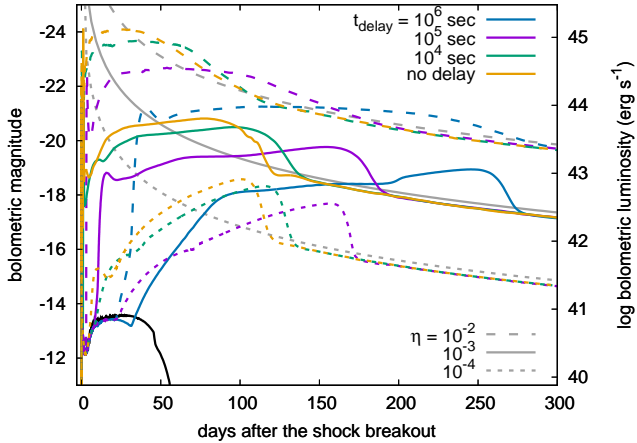}
\caption{
Bolometric LCs of fallback-powered SNe
for different $\eta$ and \tdelay. The black solid LC is from the model without fallback accretion power input. The gray lines show the accretion power injected at the inner
boundary for different $\eta$.
\label{fig:lightcurves}}
\end{figure}

Fig.~\ref{fig:lightcurves} shows representative synthetic LCs of fallback-powered SNe with different $\eta$ and \tdelay. The LC model without fallback accretion power is also shown. The LC without the fallback accretion power has a faint ($8\times 10^{40}~\mathrm{erg~s^{-1}}$) and short ($40~\mathrm{days}$) plateau phase. Although one might expect a LC that is similar to SN~1987A because of the small progenitor radius (24~\Rsun), the small explosion energy makes the initial adiabatic cooling less efficient and the early LC exhibits a plateau rather than a SN~1987A-like ``dome'' shape. Because no \Ni\ is present in the ejecta, the luminosity just drops without a tail after the plateau.

Fallback accretion power input dramatically changes the LC properties. 
The accretion power input released as thermal energy pushes the innermost layers of the ejecta and a shock wave is formed. The shock wave propagates to the surface of the slowly expanding ejecta and the entire ejecta expand much faster than the ejecta without the accretion power. The faster expansion makes the photospheric radius much larger than that in the model without the accretion power and the accretion powered LCs become much brighter. The thermal energy from the accretion keeps heating the ejecta and the extra heating keeps the accretion powered models much brighter than the model without the accretion power.
For the smallest accretion efficiency ($\eta = 10^{-4}$),
the LCs show a gradual LC rise for $100-150$~days, and their shapes resemble that of SN~1987A.
The total injected energy in these models is of the order of $10^{51}~\mathrm{erg}$ (Table~\ref{tab:energygain}). The standard explosion energy combined with the small progenitor radius results in SN~1987A-like LCs. For $\tdelay = 10^5~\mathrm{sec}$, the model only gains $3\times 10^{50}~\mathrm{erg}$, which is comparable to the original explosion energy of $2\times 10^{50}~\mathrm{erg}$.
The LCs with the larger \tdelay\ have longer rise times because of the slower expansion. The time of the LC drop determining the LC peak appears when the recombination wave in the hydrogen-rich envelope reaches at the bottom of the hydrogen-rich envelope. The less efficient adiabatic cooling caused by the slower expansion delays hydrogen recombination and the hydrogen recombination wave propagates slower. When $\tdelay = 10^{6}~\mathrm{sec}$, we find that the amount of the fallback energy is not enough to affect the LC.

\begin{deluxetable}{ccc}
\tablecaption{Total injected energy from fallback accretion \label{tab:energygain}}
\tablehead{
\colhead{$\eta$} & \colhead{\tdelay} & \colhead{$E_\mathrm{fallback}$\tablenotemark{a}} \\
 & \colhead{(sec)} & \colhead{($10^{51}~\mathrm{erg}$)} 
}
\startdata
$10^{-4}$ & 0  & 2.4 \\
      & $10^4$ & 1.4 \\
      & $10^5$ & 0.33 \\
\hline
$10^{-3}$  & 0 & 24 \\
      & $10^4$ & 14 \\
      & $10^5$ & 3.3 \\
      & $10^6$ & 0.81 \\
\hline
$10^{-2}$  & 0 & 240 \\
      & $10^4$ & 140 \\
      & $10^5$ & 33 \\
      & $10^6$ & 8.1 \\      
\enddata
\tablenotetext{a}{$E_\mathrm{fallback}=\int L_\mathrm{fallback}\,dt$.}
\end{deluxetable}

\begin{figure}
\epsscale{1.2}
\plotone{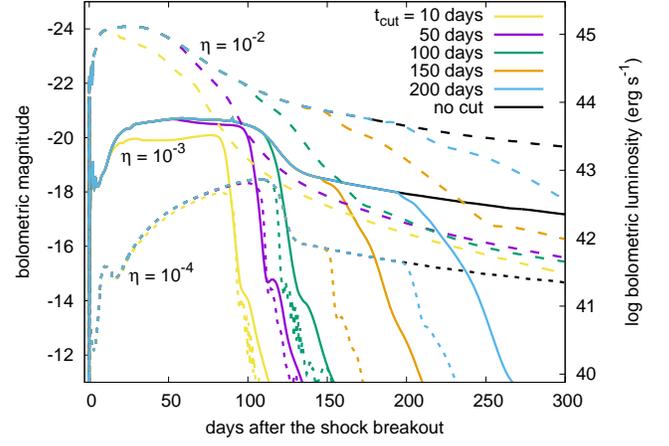}
\caption{
Bolometric LCs of the models with $\tcut=10$, $50$, $100$, $150$, and $200~\mathrm{days}$
with different accretion efficiency
$\eta$. The black LCs have $\tcut=\infty$ and are shown for comparison. All the models have $\tdelay = 0$. 
\label{fig:lightcurves_cut}}
\end{figure}

\begin{figure}
\epsscale{1.2}
\plotone{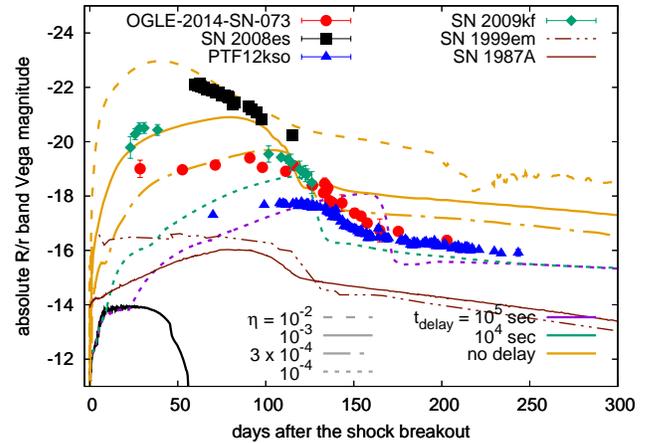}
\caption{
Comparison between the synthetic $R$-band LCs and luminous hydrogen-rich SN LCs. The dot-dashed LC model is the one with $\eta = 3\times 10^{-4}$ and $\tdelay = 0$.
See the text for the sources of the observational data. The LCs of the standard Type~IIP SN~1999em \citep{bersten2009iipbol} and SN~1987A (\citealt{arnett1989sn1987a} and the references therein) are presented for comparison.
\label{fig:lccomp}}
\end{figure}

The models with $\eta = 10^{-3}$ and $\tdelay \lesssim 10^4~\mathrm{sec}$ gain more than $10^{52}~\mathrm{erg}$ from fallback accretion power (Table~\ref{tab:energygain}). The $\tdelay = 10^5~\mathrm{sec}$ model gains $3\times 10^{51}~\mathrm{erg}$. In either case, a strong shock is generated by the fallback accretion power and provides sufficient thermal energy to sustain a luminous ($\sim 10^{43}~\mathrm{erg~s^{-1}}$) plateau for a long time ($\sim 150~\mathrm{days}$).
When $\tdelay = 10^6~\mathrm{sec}$, the LC first follows the LC without the fallback power and then the luminosity slowly keeps increasing until the LC drop caused by the hydrogen recombination. 
The models with $\eta = 10^{-2}$ gain more than $10^{53}~\mathrm{erg}$ from  fallback accretion. For
this high accretion efficiency, the bolometric LCs have a plateau-like phase of the constant luminosity, but there
is no clear drop at the phase of constant luminosity as seen in the $\eta \leq 10^{-3}$ models. The LC shapes are rather round as seen in stripped-envelope SNe.
This is because the large energy input can keep the hydrogen-rich envelope ionized and no effect of the hydrogen recombination that makes the plateau and drop in the LCs of the smaller $\eta$ appears in this case.

All the models presented so far have $\tcut = \infty$. Fig.~\ref{fig:lightcurves_cut} presents the LC models with  finite \tcut. Once energy injection is cut off, the LCs start to decline quickly because of the lack of an energy source. In the models with $\eta=10^{-2}$, the LCs drop immediately after \tcut. A similarly prompt drop after \tcut\ is also found in the models with $\eta =10^{-4}$ and $10^{-3}$ if $\tcut \geq 150~\mathrm{days}$. When accretion
power is cut during the recombination phase as in the 
models with 
$\tcut \leq 100~\mathrm{days}$ and 
$\eta = 10^{-4}$ or $10^{-3}$, however, the LC does not
drop immediately after \tcut. The remaining thermal energy keeps the recombination wave in the ejecta and the plateau phase is sustained for a while even after \tcut. The models with $\tcut = 10~\mathrm{days}$ decline at around 80~days, the
models with $\tcut = 50~\mathrm{days}$ decline at around 100~days, and the models with $\tcut = 100~\mathrm{days}$  decline at around 120~days.

The energy injection by fallback accretion may be accompanied by a significant outflow, and up to around two thirds of the accreted mass may be turned back to the ejecta \citep[e.g.,][]{kohri2005accretion}. The total accreted mass becomes 13~\Msun\ for the case of $\tdelay=0$, but the possibility that the accreted mass turns back to the ejecta is not considered in our models.  The possible turn back may make the ejecta at most about two times more massive.  When a LC has a plateau phase caused by the hydrogen recombination, as in the $\eta=10^{-3}$ models, the plateau duration would be longer by a factor of $1.4$ and the plateau luminosity would be smaller by the same factor when the ejecta mass is doubled \citep[e.g.,][]{kasen2009iip}.  If a LC is dominated by diffusion, as in the $\eta = 10^{-4}$ and $10^{-2}$ models, the LC rise time would be longer by a factor of $1.7$ and the peak luminosity would get smaller accordingly by doubling the ejecta mass \citep{arnett1982typeilc}.  Thus, our LC prediction may be altered by such a factor by taking the mass increase in the ejecta into account.

\begin{figure}
\epsscale{1.2}
\plotone{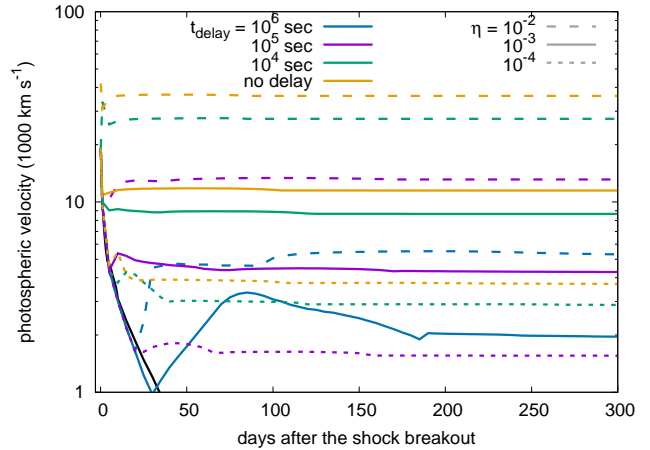}
\caption{
Photospheric velocities of the models presented in Fig.~\ref{fig:lightcurves}.
\label{fig:photov}}
\end{figure}

\begin{figure}
\epsscale{1.2}
\plotone{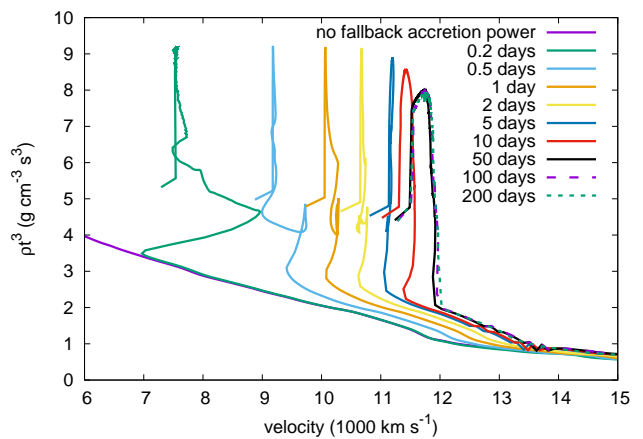}
\caption{
Inner density ($\rho$) evolution of the model with $\eta=10^{-3}$ and $t_\mathrm{delay} = 0$ in the velocity coordinate.
\label{fig:density}}
\end{figure}

In Fig.~\ref{fig:lccomp} we compare our synthetic $R$-band LCs to those of luminous hydrogen-rich SNe. As presented in \citet{moriya2018fallback}, the overall LC properties are consistent with the luminous energetic hydrogen-rich SN OGLE-2014-SN-073 \citep{terreran2017ogle73}. Some of our synthetic LCs show a similar behavior to SN~2009kf, which is among the most luminous SNe~IIP \citep{botticella2010sn2009kf}, and to PTF12kso,
which is one of the most luminous SNe~II discovered by PTF \citep{taddia2016longrisingtypeii}. Their luminosity is consistent with the LC models with $\eta = 10^{-4}-10^{-3}$. 
Our hydrogen-rich fallback-powered model with the high energy conversion efficiency ($\eta = 10^{-2}$) has a LC similar to SN~2008es, a SLSN with broad hydrogen spectroscopic features. Most hydrogen-rich SLSNe are Type~IIn but SN~2008es is one of a few SLSNe with broad hydrogen features \citep{miller2009sn2008es,gezari2009sn2008es,inserra2016slsnbroadh}.

Fig.~\ref{fig:photov} shows the photospheric velocity evolution of our models. The photospheric velocity is mostly constant in our models. This is due to the way we inject the fallback accretion energy into the ejecta. We put the fallback accretion power in the progenitor at or shortly after the shock breakout and, therefore, the initial density structure at the beginning of the energy injection is already set by the initial neutrino-driven explosion. The original hydrodynamic structure of the progenitor has already been altered by the shock and the second shock initiated by the fallback accretion energy input travels in the altered structure as presented in Fig.~\ref{fig:density}. This results in the formation of the dense shell in the ejecta
(Fig.~\ref{fig:density}), as is also found when  magnetar spin-down power is injected at the bottom of SN ejecta  \citep[e.g.,][]{kasen2010magnetar}\footnote{This dense shell is found to be unstable and deformed in multi-dimensional hydrodynamic simulations \citep[e.g.,][]{chen2014pisnmixing,suzuki2017multidcent}.}. The photosphere remains in this massive shell for a very long time and hence the photospheric velocity is kept constant for a long time. By contrast, the fallback accretion energy was turned on at 100~s after the explosion in \citet{moriya2018fallback}. This is well before the shock created by the initial explosion reaches the progenitor surface and the shock initiated by the fallback accretion energy input itself determines the initial SN ejecta structure in the models in \citet{moriya2018fallback}. No shell structure is found in the ejecta as in Fig.~\ref{fig:density} and, therefore, no constant photospheric velocity is found for a long time in the models presented in \citet{moriya2018fallback}. The photospheric velocities in SLSNe tend to decline after the LC peak, but some of them keeps the constant photospheric velocities \citep[e.g.,][]{nicholl2015slsndiversity,liu2017modjaz}. The photospheric temperature for both models is around 6,000~K at the hydrogen recombination phase. Before the hydrogen recombination, the photospheric temperature reaches $\gtrsim 10,000~\mathrm{K}$.

\section{Discussion and conclusions}\label{sec:discussion}
We have presented synthetic multi-color LCs of hydrogen-rich SNe powered by the fallback accretion. We adopt a neutrino-driven explosion model of the 40~\Msun\ zero-metallicity progenitor in which the inner part of the progenitor falls back to the central compact remnant. We estimate the fallback accretion rate based on the result of the numerical neutrino-driven explosion simulation by \citet{chan2018bh} and use it to investigate the fallback-powered SN LC properties.

We have shown that  fallback-powered SNe can have a variety of LC shapes, including those of SN~1987A-like SNe, SNe~IIP, and SLSNe. Without fallback accretion energy input, the SN is predicted to be a faint SN~IIP with a plateau magnitude of $\sim -14$~mag and a short plateau duration of around 40~days. 
In this case, one does not expect a tail phase
 because no \Ni\ is ejected in the model. If we add fallback accretion power with a small efficiency ($\eta =10^{-4}$), we find that the SN can be observed as a SN~1987A-like SN like OGLE-2014-SN-073 \citep{terreran2017ogle73,moriya2018fallback}. If the efficiency is $\eta=10^{-3}$, the fallback accretion powered SN becomes a luminous SN~IIP like SN~2009kf \citep{botticella2010sn2009kf}. If the efficiency can be as high as $\eta=10^{-2}$, the SN may be observed as a SLSN with broad hydrogen lines like SN~2008es \citep{miller2009sn2008es,gezari2009sn2008es}. Indeed, the fallback energy input in this case is similar to those found to fit SLSN LCs \citep{moriya2018fallbackslsn}.

We find that the overall LC properties of the fallback accretion powered hydrogen-rich SNe are similar to those of magnetar-powered hydrogen-rich SNe \citep{bersten2016hrichmag,sukbold2017hrichmag,orellana2018hrichmag}. Overall, these two modes of energy inputs produce similar results and it is difficult to distinguish the two central engine models solely by the LCs. One possible difference is that  fallback accretion power might be easier to shut down. The accretion towards the central compact remnant could be quenched by fallback accretion energy input,
which could push away the outer infalling shell. The LCs would then drop following the quenching of the accretion (Fig.~\ref{fig:lightcurves_cut}). By contrast, the dipole emission from the spin-down of magnetars is not likely to cease immediately, although massive magnetars may suddenly transform into a BH, and the spin-down energy may stop immediately in such cases \citep{moriya2016magbh}. Another property that is unique to magnetars is the existence of an
upper limit for the rotational energy that can be extracted from them: \citet{mazzali2014grb} argued that the energy of SNe accompanying long gamma-ray bursts is limited to around $2\times 10^{52}~\mathrm{erg}$, which is the maximum rotational energy that neutron stars of around 1.4~\Msun\ can have
and can be pumped into the ejecta by a
magnetar engine. 
In other words, if we find SNe exceeding this maximum energy that can be provided by magnetars, they would be promising fallback-powered SN candidates. We note, however, that the maximum rotational energy can change depending on the neutron star mass \citep[e.g.,][]{metzger2015magdiv}.
Looking into the spectroscopic properties could be a promising way to distinguish the two engine types because fallback accretion swallows the central part of the progenitors, and a small amount of heavy elements are likely to be ejected through the accretion disk wind, whereas most of heavy elements should be ejected in the magnetar powered SNe. In order to ascertain the observational signatures, it will also be important to better explore problems such as disk formation, the formation of disk outflows, and the termination of accretion by means of multi-dimensional simulations in the future.

We note that the fallback accretion energy is more likely to be activated in hydrogen-rich progenitors from prespective of stellar evolution. It is hard to achieve the conditions required to form magnetars in hydrogen-rich progenitors \citep[e.g.,][]{heger2005magpro} because
of efficient angular momentum transport from 
the core to the hydrogen-rich envelope, in particular if dynamo action is considered.  For fallback-powered SNe, one needs to form an accretion disk near the compact remnant to have the accretion disk wind powering the SNe. Such an accretion disk can be formed even if the initial rotational period
of the core is small because shells further out
in the envelope have higher specific angular momentum
and can easily reach Keplerian velocities before
being accreted onto the central compact object.
Even if the angular momentum in the envelope is not sufficiently large, asymmetric explosions may impart sufficient angular momentum to the outer envelope by imparting some tangential velocity onto the fallback material, as found in a more energetic fallback supernova model (Chan et al. in prep.), which is enough to make an accretion disk as it falls back.

Fallback-powered SNe may also be affected by the interaction between SN ejecta and a dense circumstellar medium (CSM) surrounding them. Although the 40~\Msun\ progenitor we studied does not experience much mass loss during its evolution, 
more massive progenitors are expected to undergo significant mass loss before the explosions due to the pulsational pair instability \citep{woosley2017ppisn}. After forming a dense CSM due to the pulsational pair instability, the progenitors are expected to collapse by forming BHs and may result in the fallback-powered SNe. Therefore, the existence of a dense CSM could be common in the fallback-powered SNe. The possibility of CSM interaction in fallback SNe needs to be studied further in the future. Difference in CSM configurations with the same central engine can result in variety of LCs \citep[e.g.,][]{fischer2018blinnikov}.

Finally, fallback-powered SNe may be common among the explosions of the first stars. It has been predicted that the first stars have a top-heavy initial mass function \citep[e.g.,][]{hirano2015popiiimf}, and many of them are likely to be massive enough to form a BH during their explosions. Pair-instability SNe from the first stars \citep{heger2002popiii} are generally considered as potential SNe that could become bright enough to be observed even 
from this early epoch of cosmic evolution
in the future transient surveys, especially in near-infrared \citep[e.g.,][]{scannapieco2005pisn}. We have shown that the fallback-powered SN from the 40~\Msun\ first star can be as luminous as pair-instability SNe. The mass range of pair-instability SNe that can be as bright as SLSNe is limited to be between around 200~\Msun\ and 250~\Msun\ \citep[e.g.,][]{kasen2011pisn}. BH-forming SNe, however, likely originate from a much wider mass ranges and may therefore be considerably more common. Thus, if sufficiently high accretion efficiencies $\eta$ can be realized, fallback-powered SNe as
investigated in our study could be frequently observed in the future high-redshift transient surveys with James Webb Space Telescope (JWST) or Wide-Field InfraRed Space Telescope (WFIRST).

\acknowledgments
TJM is supported by the Grants-in-Aid for Scientific Research of the Japan Society for the Promotion of Science (JP17H02864, JP18K13585).
This work was supported by the Australian Research Council through
ARC Future Fellowships FT160100035 (BM), Future Fellowship
FT120100363 (AH), and
by STFC grant ST/P000312/1 (BM). CC  was supported by an Australian Government Research Training Program (RTP) Scholarship. 
Sergei Blinnikov is supported by RSCF 19-12-00229 on the
development of STELLA code.
Numerical computations were in part carried out on PC cluster at Center for Computational Astrophysics, National Astronomical Observatory of Japan.
 This research was undertaken with the assistance of
resources obtained via NCMAS and ASTAC  from the National Computational Infrastructure (NCI), which
is supported by the Australian Government and was supported by
resources provided by the Pawsey Supercomputing Centre with funding
from the Australian Government and the Government of Western
Australia. 
AH has been supported, in part, by a grant from Science and Technology Commission of Shanghai Municipality (Grants No.16DZ2260200) and National Natural Science Foundation of China (Grants No.11655002).

\vspace{5mm}



\software{Kepler \citep{heger2010zmstar},
          Arepo \citep{springel2010arepo},
          STELLA \citep{blinnikov1998sn1993j,blinnikov2000sn1987a,blinnikov2006sniadeflg},
          CoCoNuT-FMT \citep{muller2010coco,muller2015coco}
          }

\bibliography{references}



\end{document}